\documentclass[aps,12pt,superscriptaddress,amsfonts,amssymb,amsmath]{revtex4}

\usepackage{graphicx}
\usepackage{epsfig}
\usepackage{makeidx}
\usepackage{epstopdf}

\begin{document}

\title{Innovation diffusion equations on correlated scale-free networks}

\author{M.L.\ Bertotti \footnote{Email address: marialetizia.bertotti@unibz.it}}
\affiliation{Free University of Bozen-Bolzano, Faculty of Science and Technology, Bolzano, Italy}
\author{J.\ Brunner \footnote{Email address: johannes.brunner@tis.bz.it}}
\affiliation{TIS Innovation Park, Bolzano, Italy}
\author{G.\ Modanese \footnote{Email address: giovanni.modanese@unibz.it}}
\affiliation{Free University of Bozen-Bolzano, Faculty of Science and Technology, Bolzano, Italy}

\linespread{0.9}

\begin{abstract}

We introduce a heterogeneous network structure into the Bass diffusion model, in order to study the diffusion times of innovation or information in networks with a scale-free structure, typical of regions where diffusion is sensitive to geographic and logistic influences (like for instance Alpine regions). We consider both the diffusion peak times of the total population and of the link classes. In the familiar trickle-down processes the adoption curve of the hubs is found to anticipate the total adoption in a predictable way. In a major departure from the standard model, we model a trickle-up process by introducing heterogeneous publicity coefficients (which can also be negative for the hubs, thus turning them into stiflers) and a stochastic term which represents the erratic generation of innovation at the periphery of the network. The results confirm the robustness of the Bass model and expand considerably its range of applicability.

\end{abstract}

\maketitle

\section{Introduction}

Like the spreading of diseases, the diffusion of innovation and information in a population can be analyzed through several mathematical models. “Compartimental” models divide the population into groups of individuals who are in certain states: sound/infected, ignorant/informed etc. The evolution in time of the populations of the various compartments is described by differential equations. 
In the basic version of the models, the population of each compartment is seen as homogeneous. A possible improvement consists in the introduction of a network structure; this has been done, among others, by Boguna et al.\ \cite{Boguna,Barrat} in the statistical networks formalism (for epidemics) and by Moreno et al.\ \cite{Moreno1,Moreno2} in the stochastic network formalism (for information). Another possible approach is to assign a network whose nodes are regarded as two-state systems, and to write master equations giving the probability that a node passes from one state to the other, as a consequence of its internal dynamics and of the state of its neighbours, like in cellular automata models. With a method of this kind, Gleeson \cite{Gle} has found exact solutions of the Bass diffusion model, which is based, in the homogeneous version, upon a linear publicity term and a quadratic imitation term. 

The Bass model has been extensively employed in marketing and social sciences \cite{Meade,Gerlach}, resulting in large empirical databases of its coefficients, measured in different situations. It has also been employed for agent-based simulations of social networks modelled on real networks \cite{WattsDodds} in order to check the effect of the so-called ``influencers''. In this work we apply to the Bass model the statistical network description which has been successfully employed for epidemics. Our main aim is to follow diffusion in time, from its beginning to its end, in particular in order to see when the diffusion peak occurs, in dependence on the exponent $\gamma$ of scale-free networks and on their assortative or disassortative correlations. We also study the temporal evolution in each link class, namely we measure the diffusion times for individuals who are more or less connected, and give quantitative estimates of the anticipation of diffusion in the most connected classes. 

Our aim is therefore different from that of epidemic models, where one is mostly interested into the initial phases of the epidemic, into the epidemic threshold and into its dependence on the model parameters. Our aims is also different from that of the Gleeson method, which allows to find phases in parameters space, and from that of Moreno et al., which measures the reliability of diffusion and the load on the network, as compared to deterministic diffusion.

In recent work using this approach \cite{CSF} we found that the hubs can serve as ``monitors'' for the adoption of others and may allow to estimate the parameters of a diffusion process at its beginning, or anyway  long before the global peak, in the case when no empirical parameters are available.
A technical difficulty encountered is the explicit construction of the matrices $P(h|k)$ appearing in the network Bass equations. ($P(h|k)$ gives the conditional probability that a node with $k$ links is connected to one with $h$ links.)
It is not sufficient for our purposes to consider the nearest-neighbour average function ${\bar k}_{nn}=\sum_h h P(h|k)$, as done in \cite{Boguna}. 

In the uncorrelated case it is straightforward to write the matrices $P(h|k)$, but in the assortative and disassortative case this has never been done before, or not for large and arbitrary dimension. We have devised and given detailed analytical proofs for a method which works especially well in the assortative case, the one actually relevant for most social networks. This allowed us in \cite{CSF} to make quantitative comparisons with the uncorrelated case (and also with the disassortative case, when applicable). 

A further improvement made possible by the network structure is to make the publicity term $p$ in the Bass equation heterogeneous.
The most straightforward way to extend the Bass model in this direction is to enhance its ``trickle-down'' character, by defining $p_i$ coefficients which are larger for the hubs, while still giving the some total expenditure for publicity.

Sometimes, however, innovation proceeds from the periphery of a system, where it is generated and grows in small niches, until it becomes the ``disruptive'' innovation described by Christensen \cite{Chri}. In that case, the innovation is not launched through a marketing campaign, but starts as a stochastic and erratic process; then diffusion can accelerate or almost come to a halt several times at random, before it reaches some points in the network where it can propagate more vigorously. Empirical evidence \cite{Baptista} shows that this is often the case for hierarchical inter-firm structures like those common in Alpine regions for geographical and logistic reasons \cite{Inter}. This ``trickle-up'' process is the main subject of this work and can be modelled through $p_i$ coefficients which are larger for the less connected nodes of the network.

One can also take into account the possibility that some of the hubs are ``conservative'' and act as bottlenecks for diffusion, instead of facilitating it. This can be modelled through negative publicity coefficients, leading to a major departure from the familiar Bass model, where one has to careful adapt the equations in such a way to allow temporary negative values of the adoption rate $f$, but avoid negative values of the cumulative adoption $F$, which would be meaningless.

Finally, since our model allows to follow the detailed time evolution of the system through the numerical solution of the $N$ differential equations, we can add a stochastic term and study the resulting Langevin equations, in particular in the case of ``trickle-up'' innovation \cite{Abr}.

The outline of the article is the following. In Sect.\ \ref{structure} we recall the network Bass equations introduced in \cite{CSF} and the main features of its numerical solutions.  In Sect.\ \ref{t-up} we introduce the new trickle-up equations and give some numerical solutions.  In Sect.\ \ref{stoc} we introduce the stochastic trickle-up equations.  Sect.\ \ref{disc} contains a general discussion of the possible applications of the model and our conclusions.

\section{Network structure and trickle-down}
\label{structure}

The network Bass equation introduced in \cite{CSF} is a system of non-linear first order differential equations, in which the imitation term of the Bass model has been split over $N$ connectivity classes:
\begin{equation}
\frac{{d{G_i}(t)}}{{dt}} = [1 - {G_i}(t)]\left[ {p + iq\sum\limits_{h = 1}^N {P(h|i){G_h}(t)} } \right] \qquad i=1,...,N.
\label{BassNet}
\end{equation}
$G_i(t)=F_i(t)/P(i)$ is the fraction of potential adopters with $i$ links that at the time $t$ have actually adopted the innovation.
The matrices $P(i)=c/i^\gamma$ (link density) and $P(h|i)$ (link correlations) must obey the Network Closure Condition (NCC; see \cite{NCC}). 
Further conditions are normalization $\sum\limits_{k = 1}^N P(k|i) =1$,  $\sum\limits_{k = 1}^N P(k) =1$ and positivity $P(i|k) \geq 0$, $P(i) \geq 0$. 

\subsection{Summary of trickle-down results}

A summary of the numerical results obtained in \cite{CSF} for the trickle-down case is given in Figs.\ 1, 2 and Tab.\ I. Tab.\ II summarizes results for a wider range of the exponent $\gamma$ and includes for comparison results for the trickle-up case, to be discussed in the next section.

\begin{table}[h]
\begin{center}
\begin{tabular}{|c|c|c|c|c|c|}
\hline
  & $\gamma=2.0$   & \ $\gamma=2.25$  \ & \ $\gamma=2.5$ \ & \ $\gamma=2.75$ \ &  \ $\gamma=3.0$ \   \nonumber \\
\hline
$T$-Assort. & \ 2.5  \ & \ 4.0 \ & \  4.8 \ &  \ 5.1 \  & \ 5.4 \    \nonumber \\
\hline
$T_{50}$-Assort. & \ 0.80 \ & \ 0.64 \ &  \ 0.51 \  &  \ 0.49 \  & \ 0.49 \   \nonumber \\
\hline
$T$-Uncorr.& \ 2.4 \ & \ 2.8 \ & \ 3.5 \  &  \ 4.2 \  &  \ 4.7 \   \nonumber \\
\hline
$T_{50}$-Uncorr.& \ 1.4 \ & \  1.3 \ & \ 1.3 \   & \ 1.3 \  & \ 1.3 \  \nonumber \\
\hline
$T$-Uncorr., var. $p$ & \ 1.6 \ & \ 2.5 \ & \ 3.6 \   &  \ 4.8 \  &  \ 5.9 \  \nonumber \\
\hline

\end{tabular}
\end{center}
\caption
{Dependence on the scale-free exponent $\gamma$ of the total diffusion peak time $T$ and of the partial diffusion peak time $T_{50}$ (for individuals with 50 links) in a network with largest degree $N=100$, in the uncorrelated and assortative case. The peak time $T_{50}$ has been chosen as indicator of the anticipated diffusion in the hubs, instead of $T_{100}$, because in these networks the hubs with 100 links are a very small fraction of the total population, and thus are not significant for statistical monitoring. Nevertheless, the anticipation effect is clearly very strong, especially for the assortative networks: for $\gamma=2$, $T_{50}$ is approx. 20\% of $T$, and for $\gamma=3$ it is approx. 10 \% of $T$. Note that for uncorrelated networks the total diffusion time is smaller, but the anticipation effect is weaker. The dependence $T(\gamma)$ in the case of variable $p$ coefficients $p_k=c_1 k^\gamma p$ is also shown.}
\label{tab2}
\end{table}

All data show an anticipation effect of diffusion in the hubs, more or less pronounced, depending on several factors (maximum number $N$ of links, nature of the correlations). The procedure for the explicit construction of assortative and disassortative correlation matrices is described in \cite{CSF}.

\begin{table}[h]
\begin{center}
\begin{tabular}{|c|c|c|c|c|c|c|c|c|}
\hline
$\gamma$  & \ $T$  \ & \ $T_{N}$  \ & \ $T$ \ & \ $T$ \ &  \ $T$ \  & \ $T_N$ \  & \ $T$ \ & \ $T$ \ \nonumber \\
& (uncorr.)  & (uncorr.) & (assort.) & (disass.) & (t-down) & (t-down) & (follower\ t-up) & (stifler\ t-up) \nonumber \\
\hline
\ 1/4 \ & \ 5.1  \ & \ 4.4 \ & \  & &  \ 4.9 \ &  \ 4.1 \  &  &  \nonumber \\
\hline
\ 1/2 \ & \ 4.9 \ & \ 4.1 \ & \   & &  \ 4.5 \  &  \ 3.6 \  & \ 5.6 \ & \ 10.5 \  \nonumber \\
\hline
\ 3/4 \ & \ 4.6 \ & \ 3.8 \ & \ 4.3 \  & & \ 4.1 \  &  \ 3.1 \   &  \ 5.3 \ & \ 8.1 \ \nonumber \\
\hline
\ 1 \ & \ 4.4 \ & \  3.5 \ & \ 3.9 \   & \ 5.2 \  & \ 3.7 \ & \ 2.5 \   &  \ 5.0 \ & \ 6.8 \ \nonumber \\
\hline
\ 3/2 \ & \ 4.1 \ & \ 2.9 \ & \ 3.3 \   & & \ 3.1 \  &  \ 1.2 \  &  \ 4.5 \ & \ 5.3 \ \nonumber \\
\hline
\ 2 \ & \ 4.1 \ & \  2.5 \ & \ 3.2 \     & &  \ 3.1 \ &  no peak  & \ 4.3 \ & \ 4.7 \ \nonumber \\
\hline
\ 5/2 \ & \ 4.5 \ & \ 2.4 \ & \   & & &  & & \nonumber \\
\hline
\ 3 \ & \ 5.0 \ & \ 2.3 \ & \   & & &  & & \nonumber \\
\hline
\end{tabular}
\end{center}
\caption
{Time of adoption peak for different values of the scale-free exponent $\gamma$ in the following cases: (1) uncorrelated networks; (2) assortative/disassortative networks; (3) enhanced trickle-down diffusion, with $p_i \propto i^\gamma$; (4) trickle-up diffusion with follower hubs; (5) trickle-up diffusion with stifler hubs. For the cases (1), (3), the peak time $T_N$ for the most connected individuals is also given. In the cases (4), (5), (6) the network is taken to be uncorrelated, but it is straightforward to introduce A/D correlations. The homogeneous Bass function with the same $q$ and $p$ ($q=0.4$, $p=0.03$) peaks approx.\ at $t=6.1$.}
\label{tab1}
\end{table}

\begin{figure}
\begin{center}
\includegraphics[width=10cm,height=5cm]{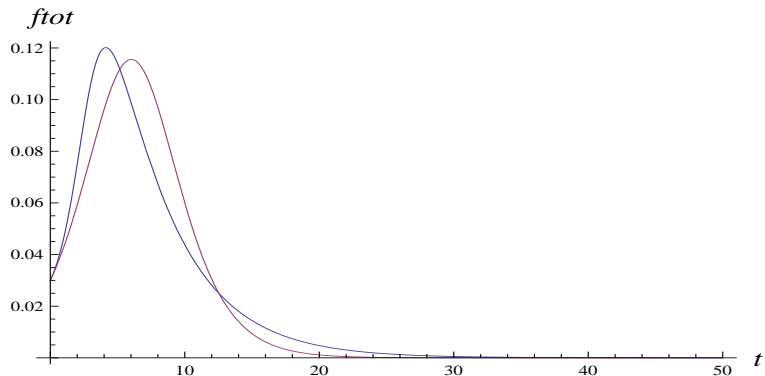}
\caption{Total adoptions $f_{tot}$ in time (blue curve), for an uncorrelated scale-free network with density proportional to $1/i$ ($\gamma=1$), with $N=15$. The violet curve is the simple Bass function with the same $q$ and $p$ ($q=0.4$, $p=0.03$). The function $f_{tot}$ peaks approx.\ at $t=4.4$, while the simple Bass function peaks approx.\ at $t=6.1$.} 
\label{f01}
\end{center}  
\end{figure}

\begin{figure}
\begin{center}
\includegraphics[width=10cm,height=5cm]{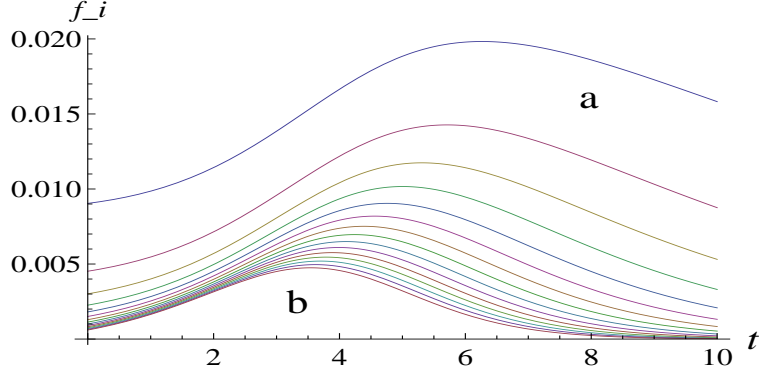}
\caption{Partial adoptions in time in the different link classes for the same parameters as in Fig.\ \ref{f01}. The function (a) which has the largest value at $t=0$ is $f_1$, representing the adoptions of individuals with only 1 link. The function $f_{15}$, which represents the adoptions of the most connected individuals (b), peaks approx.\ at $t=3.5$.} 
\label{f02}
\end{center}  
\end{figure}

\begin{figure}
\begin{center}
\includegraphics[width=10cm,height=5cm]{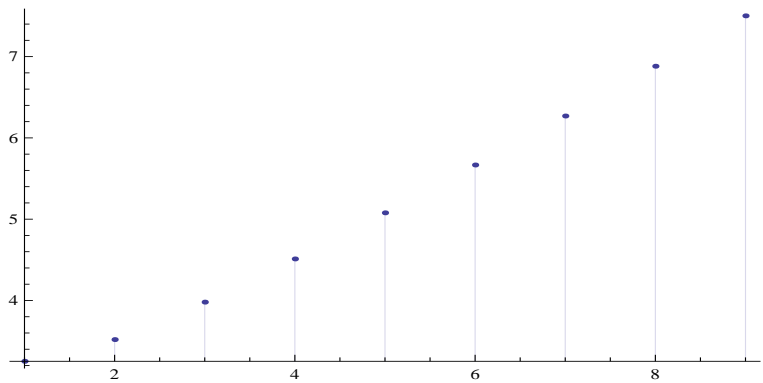}
\caption{The function ${\bar k}_{nn}$ for the assortative matrix in the example of eq.\ (\ref{example}), after normalization.} 
\label{knn}
\end{center}  
\end{figure}

An example of an assortative correlation matrix with dimension 9 and $\gamma=3/4$, $\alpha=1$ before normalization is the following: 
\begin{equation}
\left(
\begin{array}{ccccccccc}
 1 & 1 & \frac{1}{2} & \frac{1}{3} & \frac{1}{4} & \frac{1}{5} & \frac{1}{6} & \frac{1}{7} & \frac{1}{8} \\
 \sqrt[4]{2} & 1 & 1 & \frac{1}{2} & \frac{1}{3} & \frac{1}{4} & \frac{1}{5} & \frac{1}{6} & \frac{1}{7} \\
 \frac{\sqrt[4]{3}}{2} & \sqrt[4]{\frac{3}{2}} & 1 & 1 & \frac{1}{2} & \frac{1}{3} & \frac{1}{4} & \frac{1}{5} & \frac{1}{6} \\
 \frac{\sqrt{2}}{3} & \frac{1}{2^{3/4}} & \frac{\sqrt{2}}{\sqrt[4]{3}} & 1 & 1 & \frac{1}{2} & \frac{1}{3} & \frac{1}{4} & \frac{1}{5} \\
 \frac{\sqrt[4]{5}}{4} & \frac{\sqrt[4]{\frac{5}{2}}}{3} & \frac{\sqrt[4]{\frac{5}{3}}}{2} & \frac{\sqrt[4]{5}}{\sqrt{2}} & 1 & 1 & \frac{1}{2} & \frac{1}{3} & \frac{1}{4} \\
 \frac{\sqrt[4]{6}}{5} & \frac{\sqrt[4]{3}}{4} & \frac{\sqrt[4]{2}}{3} & \frac{\sqrt[4]{\frac{3}{2}}}{2} & \sqrt[4]{\frac{6}{5}} & 1 & 1 & \frac{1}{2} & \frac{1}{3} \\
 \frac{\sqrt[4]{7}}{6} & \frac{\sqrt[4]{\frac{7}{2}}}{5} & \frac{\sqrt[4]{\frac{7}{3}}}{4} & \frac{\sqrt[4]{7}}{3 \sqrt{2}} & \frac{\sqrt[4]{\frac{7}{5}}}{2} & \sqrt[4]{\frac{7}{6}} & 1 & 1 & \frac{1}{2} \\
 \frac{2^{3/4}}{7} & \frac{1}{3 \sqrt{2}} & \frac{2^{3/4}}{5 \sqrt[4]{3}} & \frac{1}{2\ 2^{3/4}} & \frac{2^{3/4}}{3 \sqrt[4]{5}} & \frac{1}{\sqrt{2} \sqrt[4]{3}} & \frac{2^{3/4}}{\sqrt[4]{7}} & 1 & 1 \\
 \frac{\sqrt{3}}{8} & \frac{\sqrt{3}}{7 \sqrt[4]{2}} & \frac{1}{2\ 3^{3/4}} & \frac{\sqrt{\frac{3}{2}}}{5} & \frac{\sqrt{3}}{4 \sqrt[4]{5}} & \frac{1}{\sqrt[4]{2} 3^{3/4}} & \frac{\sqrt{3}}{2 \sqrt[4]{7}} & \frac{\sqrt{3}}{2^{3/4}} & 1 \\
\end{array}
\right).
\label{example}
\end{equation}
The corresponding strictly increasing ${\bar k}_{nn}$ function after normalization is shown in Fig.\ \ref{knn}.

In view of the crucial role of the hubs in the diffusion process it is clear that, fixed a certain total amount available for publicity, it is more effective to address the advertisement to them. This suggests a generalization to variable $p$ coefficients. Since the individuals with $k$ links are $k^\gamma$ times less numerous than those with one link, if the $p$ coefficient (supposed proportional to the expenditure for publicity) is $k^\gamma$  times larger for the former than for the latter, then the total cost of publicity will be the same. Denoting by $p_k$ the publicity coefficient for the link class $k$ this leads to write the equations (\ref{BassNet}) as
\begin{equation}
\frac{{d{G_i}(t)}}{{dt}} = [1 - {G_i}(t)]\left[ {p_i + iq\sum\limits_{h = 1}^N {P(h|i){G_h}(t)} } \right] \qquad i=1,...,N,
\label{BassNet-pvar}
\end{equation}
where we set $p_k=c_1 k^\gamma p$, being $c_1$ an appropriate normalization constant determined through the condition $\sum_k p_k P(k)=p$, whence $c_1=1/(cN)$.
The numerical solutions of these equations show that diffusion indeed becomes still faster, both for the total population and for the hubs (Tab.\ I).

\section{Trickle-up equations}
\label{t-up}

\begin{figure}
\begin{center}
\includegraphics[width=10cm,height=5cm]{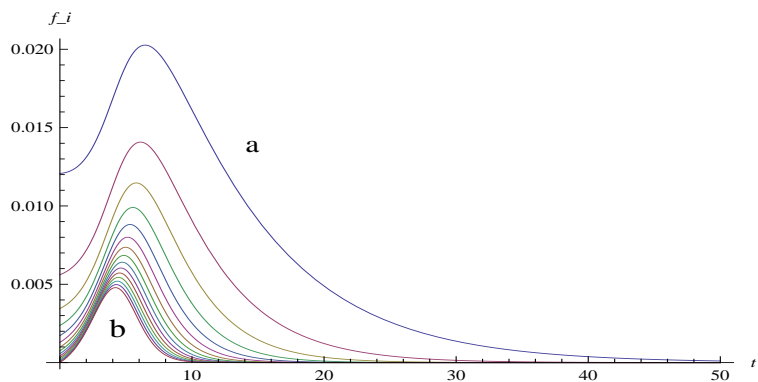}
\caption{Partial adoption in time in the case of trickle-up diffusion with follower hubs ($\gamma =1$). Adoption from the hubs is only slightly anticipated (compare Fig.\  \ref{f02}). Curve (a): individuals with 1 link; (b): individuals with 15 links.} 
\label{follow-t-up}
\end{center}  
\end{figure}

We are now going to employ the system of differential equations (\ref{BassNet-pvar}) for the description of a trickle-up diffusion process on scale-free networks with variable exponent $\gamma$. Our aim is, as before, to obtain the partial and total diffusion curves and to compute the peak times; and more generally, to compare the trickle-down and trickle-up process. While in the trickle-down case the publicity terms $p_i$ were directly proportional to a power of the number of links ($p_i \propto i^\gamma$), for the trickle-up we consider terms with $p_i$ which decreases linearly in $i$. Accordingly, they take their maximum value for the least connected individuals and are small or zero for the hubs. Such terms do not represent publicity anymore, but an effect of ``grassroots creation'' of the innovation. A possible expression is
\begin{equation}
p_i^{(f)} = \left( 1-\frac{i}{N} \right) c_2 p, \qquad \qquad {\rm (follower \ hubs)}
\label{pi1}
\end{equation}
where $c_2$ is a normalization constant such that $\sum_i p_i P(i)=p$. From initial conditions $G_i(0)=0$, the partial adoption functions $G_i$ will therefore grow more quickly for small $i$, representing a diffusion process which starts at the periphery of the network and propagates towards the hubs, whence it can further diffuse thanks to the imitation terms. 

One observes (Tab.\ \ref{tab1}) that diffusion is on the whole slower and, as expected, adoption from the hubs does not anticipate adoption at the periphery, but occurs approximately at the same time (Fig.\ \ref{follow-t-up}). This may be compared with the empirical and theoretical results of \cite{Goldenberg}, where hubs of this kind are called ``follower hubs'', in order to distinguish them from innovator hubs.

\begin{figure}
\begin{center}
\includegraphics[width=10cm,height=5cm]{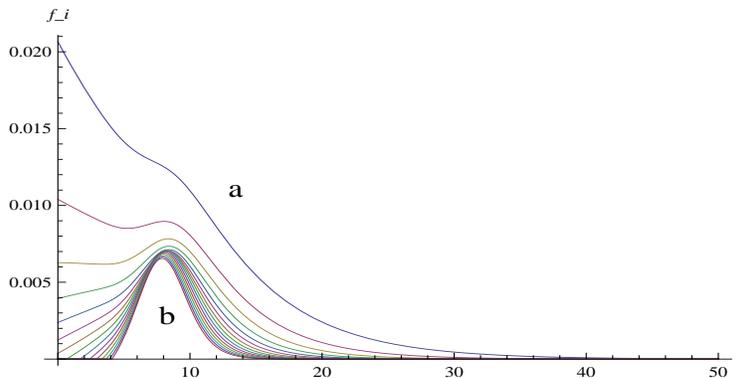}
\caption{Partial adoption in time in the case of trickle-up diffusion with stifler hubs with negative $p_i$ coefficients ($\gamma =3/4$). The adoption rate of the hubs remains zero until the imitation term exceeds the absolute value of their $p_i$ and forces adoption. Note that the adoption rate of the least connected individuals is decreasing also at small time. Curve (a): individuals with 1 link; (b): individuals with 15 links.} 
\label{t-up-stifler}
\end{center}  
\end{figure}

The next step is to consider situations where innovation is born in the periphery and conservative, or``stifler" hubs try to stop it, in the sense that they have a negative $p$ coefficient and adopt only when their imitation term exceeds the absolute value of $p$. We suppose $p$ to be the more negative, the larger is the number of links, for instance according to the function
\begin{equation}
p_i^{(s)} = \left( 1-\frac{2i}{N} \right) c_3 p, \qquad \qquad {\rm (stifler \ hubs)}
\end{equation}
where, again, $c_3$ is a normalization constant such that $\sum_i p_i P(i)=p$; note that $c_3$ is positive, as long as $\gamma>0$.
The introduction of negative $p$ coefficients is a major departure from the original Bass model. If the negative-$p$ terms are not properly controlled, they can lead to a behavior of the equations inconsistent with the model, like the diminution of the total adoptions in certain time intervals (a de-adoption phenomenon which can exist in reality, but is not considered in the model), or even to negative values of $G$, which are meaningless. We shall therefore modify the equations adding a suitable $\theta$-function which sets the time derivative of $G_i$ to zero when the negative-$p$ term dominates the imitation term: 
\begin{equation}
\frac{{d{G_i}(t)}}{{dt}} = [1 - {G_i}(t)]\left[ {p_i^{(s)} + iq\sum\limits_{h = 1}^N {P(h|i){G_h}(t)} } \right] \theta \left( {p_i^{(s)} + iq\sum\limits_{h = 1}^N {P(h|i){G_h}(t)} } \right) \qquad i=1,...,N.
\label{BassNet-pvar2}
\end{equation}
So in some cases adoption is blocked in a link class, but not reversed, until imitation of the other link classes prevails. A typical solution is shown in Fig.\ \ref{t-up-stifler}. The total adoption peak times are markedly larger (Tab.\ \ref{tab1}).

\section{Stochastic equations for trickle-up}
\label{stoc}

In this section we analyze the stochastic, erratic aspect of the generation of innovation through trickle-up from the network periphery, as discussed in the Introduction. For this purpose we write a suitable $N$-dimensional Langevin equation \cite{Carro} comprising a deterministic term and a stochastic diffusion term, namely 
\begin{equation}
dG_i(t)=[1 - {G_i}(t)]\left[ {p_i^{(f)} + iq\sum\limits_{h = 1}^N {P(h|i){G_h}(t)} } \right]dt +
[1-G_i(t)]p_i^{(f)}\Gamma \eta_i \sqrt{dt} ,
\end{equation}
where $\Gamma$ is the noise amplitude and $\eta_i$ are random Gaussian variables with zero average and $\sigma=1$. Note that the publicity term is such to define follower hubs.

A detailed statistical analysis of the results of the Langevin equation will be presented elsewhere. It turns out, however, that although the diffusion curves of the single link classes can be quite affected by the noise, the networking effect on diffusion produces a total adoption curve which is quite smooth, and hardly distinguishable from a deterministic Bass-like curve (except for some fluctuations in the initial phase). For instance, with a scale-free exponent $\gamma=1$, the deterministic trickle-up process with follower hubs has a total peak diffusion time $T=5.0$ (Tab.\ \ref{tab1}); with noise amplitude $\Gamma=1$, i.e.\ noise-to-signal ratio approximately equal to 1, we obtain variable peak times, but the standard deviation of their average is small: $T=5.0\pm 0.1$. With $\Gamma=2$ the total adoption curve is still quite smooth and we obtain $T=5.0\pm 0.2$. 

This highlights, in our opinion, one of the reasons why the Bass model is so successful, in spite of its simplicity: it is robust with respect to inhomogeneous noise. In other words, uncorrelated random variations of the diffusion rate occurring in different sectors of the population tend to average out, especially when these variations are small in the hubs, like in a trickle-up process. 

\section{Discussion and conclusions}
\label{disc}

The initial inspiration for this work came from some preliminary results of an analysis of inter-firm innovation networks in the alpine region of South Tyrol. This analysis confirmed the importance of network connections in the spreading of innovations, as already reported by other studies, and suggested that the network structure should be explicitly inserted into one of the models most widely employed for the description of innovation diffusion, namely the Bass equation. Our experience with the local network structures also pointed to the importance of trickle-up innovation process, which are absent from the traditional Bass model and have been rarely studied in the literature. Actually, a trickle-up process can be only simulated in a model with a network structure, so we saw here a chance to improve the Bass equation under the two respects at once.

We have successfully integrated the network structure into the equations of the original model and we have studied in particular the total diffusion time and the partial diffusion times (in the link classes) in dependence on the model parameters. This was done separately in the cases of diffusion originating uniformly in the network, or mainly in the hubs, or mainly at the periphery. Further technical improvements have been the explicit construction of correlation matrices and suitable modifications of the differential equations in order to allow negative linear terms (representing stifler hubs) or stochastic terms (representing random generation of innovation).

Results come in the form of accurate time diffusion curves and of numerical values assessing the anticipation effect in the hubs; for instance the data in Table II show that for assortative networks the anticipation effect is strongest and grows with the network exponent $\gamma$, while for uncorrelated networks the total diffusion time is smaller, but the anticipation effect is weaker. These results show that even for the traditional trickle-down Bass model, the introduction of the network offers the following advantages:

1) At the purely theoretical level of the numerical simulation of diffusion, one obtains a very efficient scheme in which all parameters can be varied and solutions obtained essentially in real time (as compared, for instance, to agent-based simulations). This allows to explore numerically the predicted effects of the network structure, of varying scale-free exponents, correlations etc.

2) Suppose we want to apply the Bass equation to a given ``market'' and to a given product for which the $p$ and $q$ parameters have already been estimated (for instance a new home appliance or electronic device falling into some known category), and we want to improve the accuracy of our predictions by considering the local network structure. Our method then indicates which informations on the network are needed, and how they could be collected. For instance, one could organize interviews with a sample of the population in which each respondent is first asked to recall a previous adoption of a similar product and to estimate the number of persons to which he personally showed the new adoption; this would allow an estimate of the link density $P(k)$. A further, more sophisticated step would be to have each respondent in the sample listing the names of the individuals involved in those communications related to the adoption; this allows to compute a posteriori the correlations in the sample. Note that we cannot address here in detail the issue of the statistical significance of these data, size of the sample needed in relation to statistical errors etc.

3) For consumer products whose $p$ and $q$ coefficients cannot be estimated by analogy with other products, our network Bass model suggests possible ways to estimate the parameters while diffusion is unfolding but is still far from its total maximum adoption rate. For this purpose, one should monitor the adoption of a certain number of hubs, previously identified as explained above, and fit their adoption curve by an equation system with the appropriate exponent $\gamma$. Also for this kind of applications, a detailed statistical method including the handling of estimation uncertainties must be developed. We would like to stress, however, that our network version of the Bass model, offering this possibility, is a clear progress in comparison to other refinements of the model which take into account multiple generations of products, multiple markets etc. A similar enhancement can also be applied to other diffusion models, like those for epidemics; a crucial feature for the detailed study of time evolution would be also in those cases the explicit construction of suitable correlation matrices, while Ref.\ \cite{Boguna} mainly focuses on asymptotic analysis (e.g., on the epidemic threshold).

Concerning the trickle-up version of the network Bass equation, the best set of real data presently available for a comparison with our model is reported by Goldenberg et al. \cite{Goldenberg}
who have analyzed the diffusion of some online services in a large Korean social network with $22$ million users. 
After identifying the network hubs and finding a scale-free degree distribution, the authors of \cite{Goldenberg} measured several quantities 
related to the total speed of the adoption process, to hub and non-hub adoption timing and to hub adoption as predictor of total adoption timing.
The results are generally in agreement with the dynamics predicted by our model. 
A simple regression analysis and comparison with the homogeneous Bass model is also given in \cite{Goldenberg}. 
The main focus of that work, however, is on marketing aspects.

\end{document}